\documentclass[12pt]{article}
\usepackage[T1]{fontenc}
\usepackage[utf8]{inputenc}
\usepackage[letterpaper, margin=1.0in]{geometry}
\usepackage[square, authoryear]{natbib}
\usepackage{rotating,graphicx}
\usepackage{amsmath}
\usepackage{amssymb}
\usepackage{amsthm}
\usepackage{mathtools}
\usepackage{bm}
\usepackage{color}
\usepackage{caption}
\usepackage{subcaption}
\usepackage{lineno}
\usepackage{hyperref}
\usepackage{cleveref}
\usepackage{authblk}
\usepackage{appendix}
\usepackage{setspace}

\newcommand{\ve}[1]{\mathbf{#1}}
\newcommand{\eps}{\epsilon}
\newcommand{\Req}{r_{\eps}}

\newcommand{\del}{\delta}
\newcommand{\dd}{\mathrm{d}}
\newcommand{\Tri}{\mathcal{T}}
\newcommand{\normal}{\mathbf n}
\newcommand{\obs}{\mathbf{x}}

\newcommand{\Du}{\Delta\mathbf{u}}

\title{Three dimensional non-singular mollified elastic dislocation theory for extended width fault zones and inhomogeneous boundary element models}
\author[1]{Brendan Meade}
\affil[1]{Department of Earth \& Planetary Sciences, Harvard University, Cambridge, Massachusetts, meade@fas.harvard.edu}
\date{}

\begin{document}
\maketitle

\clearpage
\section*{Key points}

\begin{itemize}
    \item Closed form integration of extended-width elastic displacement-discontinuity kernels over arbitrary triangular elements
    \item On-fault elastic stress becomes physically interpretable once the anelastic eigenstress of the distributed slip is subtracted
    \item Bounded kernels allow for well-posed collocation boundary element models with material property variations and non-planar topography
 \end{itemize}

\clearpage
\begin{abstract}
Classical elastic dislocation theory (CEDT) has two challenges when applied to faulting problems: 1) fictitious on-fault stresses not defined by ordinary integration and 2) the geometric unreality of infinitely thin fault zones.  We show that both can be resolved by a mollified elastic dislocation theory (MEDT) built on Cortez blob \citep{Cortez2001} mollified displacement discontinuity Green's functions, which represent deformation across spatially distributed fault zones of finite scale $\eps$ and produce singularity-free displacements and stresses everywhere.  Analytical integration of the mollified source solution over arbitrary planar triangular elements is done with AI, and the resulting closed-form solutions allow for the calculation of non-singular stresses across geometrically complex fault systems.  Further, we demonstrate the decoupling of the fault-zone width scale $\eps$ from the mesh length scale $h$, the elasticity analog of a result established for regularized viscous flow Stokeslets \citep{FerrantiCortez2024}.  We use these mollified kernels to demonstrate numerically stable collocation boundary element models including spatially extended fault zones, material property variations and non-planar topography.
\end{abstract}

\clearpage
\section*{Plain language summary}
The Earth's crust deforms elastically in response to fault slip. A classical approach to describing the static displacements and stresses generated by earthquakes has two long-standing problems.  First, it predicts unphysical, and in select locations infinite, on-fault stresses.  Second, it treats faults as infinitely thin cuts, while observations suggest that real fault zones have extended widths.  We show how both problems can be addressed by distributing fault slip across an adjustable width zone rather than concentrating it on an infinitely thin surface. We describe exact formulas for this finite-stress approach, for geometrically complex fault surfaces. Because the formulas are exact, the zone width can be chosen freely, independent of how finely the fault is divided.  We show how this approach can be used to compute the displacements resulting from fault slip in models with topography and material property variations.

\clearpage
\section{Introduction}
\begin{figure}[htp]
  \centerline{\includegraphics[width=0.60\textwidth]{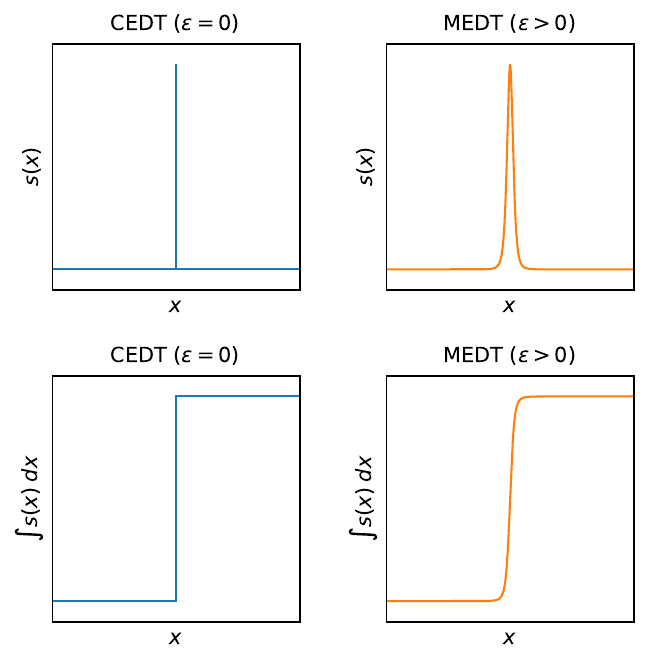}}
    \caption{Cartoon of kernel point (left column) and Cortez blob (right column) kernel shapes (top row) and cumulative ``slip'' across a fault-like structure for each of these cases (bottom row).}
    \label{fig:fig_blob_slip_cartoon}
\end{figure}

Classical elastic dislocation theory (CEDT), the representation of faulting as displacement discontinuities across surfaces embedded in a linear elastic medium \citep[e.g.,][]{Volterra1907, Steketee1958, ComninouDundurs1975, Okada1985, Okada1992, JeyakumaranRudnickiKeer1992, NikkhooWalter2015}, is widely used in earth science to represent the quasi-static crustal response to fault slip \citep[e.g.,][]{Okada1985, KingSteinLin1994, SimonsFialkoRivera2002, Segall2010}.  CEDT also carries two conceptual challenges.  The first is that on-fault stresses are not defined by ordinary integration because the stress kernel is hypersingular, so the theory fails to provide the meaningful on-fault stresses that fault mechanics most needs \citep{Eringen1977, Eringen1983}.  The second is that real faults accommodate slip across zones of finite width, from principal slip zones centimeters or less thick within fault cores and damage zones of order 100~m \citep[e.g.,][]{ChesterEvansBiegel1993, Sibson2003, FaulknerEtAl2010} to zones of distributed coseismic surface deformation tens to hundreds of meters wide resolved geodetically \citep[e.g.,][]{Milliner2015, Barnhart2020, Li2023}, while CEDT concentrates slip on an infinitely thin representation of a fault surface.  This paper describes how both challenges can be addressed through the development of mollified elastic dislocation theory (MEDT), using a specific blob mollification \citep{Cortez2001} to create Green's functions for the displacements, strains, and stresses associated with spatially distributed slip zones characterized by a length scale $\eps$ (Figure \ref{fig:fig_blob_slip_cartoon}) which can be interpreted either as a tunable numerical smoothing scale or as a physical scale for the effective width of a real-world fault zone. This fundamental solution can be analytically integrated over individual triangular elements allowing the construction of non-singular stress fields on arbitrary triangulated representations of fault surfaces.

\begin{figure}[htp]
    \centerline{\includegraphics[width=1.00\textwidth]{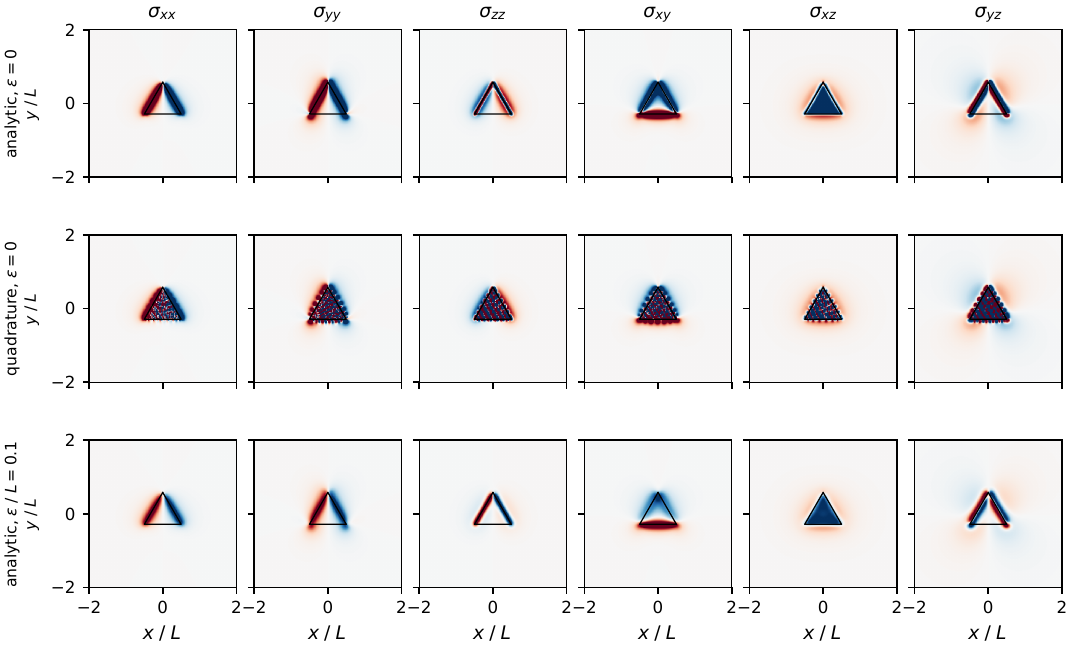}}
    \caption{Six stress components on a horizontal evaluation plane $z=0.05L$ above a unit-edge equilateral triangle with right-lateral strike-slip. Top: analytic integration of the singular kernel ($\eps=0$). Middle: Gauss quadrature of the singular kernel ($\eps=0$, $n_q=8$ per side). Bottom: analytic integration of the mollified kernel ($\eps/L=0.1$).}
    \label{fig:fig_triangle_stress_components}
\end{figure}

The fundamental problem is that the point source solutions that map a displacement discontinuity (DD, fault-slip) to stress, $\sigma_{mn}(\obs) = \int_\Tri K_{mn,k}(\ve g)\,\Du_k\,\dd S$, are hypersingular \citep[e.g.,][]{Bonnet1995, sladek1998singular, Aliabadi2002},

\begin{equation}
  K = \frac{\mu f(\mathbf g)}{4\pi(1-\nu)r^{3}}
\end{equation}

where $\mu$ is a shear modulus, $\nu$ is Poisson's ratio, $f(\mathbf g)$ is a dimensionless function of triangle and observer geometry, and $r$ is the distance between the observation and displacement discontinuity source points.  Gaussian quadrature converges for observation points at some distance from the source element, but fails when the observation distance is comparable to or less than the element size (Figure \ref{fig:fig_triangle_stress_components}), which is exactly the case relevant to calculating the on-fault stresses required for both fault mechanics modeling and matrix assembly for collocation boundary element models \citep[e.g.,][]{martin1989hypersingular, krishnasamy1992continuity, GuiggianiM.1992AGAf}.

In the face of this challenge a range of insightful approaches has been developed, spanning analytical regularization, higher-order polynomial representations of fault slip, and methods that avoid the classical singularity.  An array of regularization/smearing approaches \citep[e.g.,][]{Eringen1977, GutkinAifantis1999, LazarMaugin2005, CaiArsenlisWeinberger2006, LotheHirth2005} replace the singular terms with a $r \to \sqrt{r^2+\epsilon^2}$-like substitution. These are very similar to the approach used here, though the resulting integrands often lack closed-form integrals over the elements necessary to tessellate geometrically complex fault surfaces and so reintroduce numerical quadrature, making them insufficient for general collocation boundary element models. More physically motivated approaches to the singularities on crack/dislocation surfaces have introduced along-surface slip gradations and near-tip cohesive zones \citep[e.g.,][]{Peierls1940, Nabarro1947, Dugdale1960, Barenblatt1962, Ida1972, Rice1992}. Again, the mollified kernel approach used here is closely related to many prior formulations with two notable differences: 1) the method here is applied over an entire fault rather than near the tip and 2) the smoothing length scale appears analytically in a new kernel instead of through constitutive relationships for near-tip behavior. Another approach is largely focused on developing analytic higher order (typically linear or quadratic-variation) displacement discontinuity elements \citep[e.g.,][]{CrawfordCurran1982, ShouCrouch1995, ShouNapierCrouch1997}. These methods have been applied to planar problems with non-constant slip.  Quadrature by expansion \citep{KlocknerBarnettGreengard2013} is a numerical approach that works by extrapolating non-singular field values towards locations where the solution would be analytically singular. This approach requires reasoned choices of expansion radius and quadrature order, both linked implicitly to the mesh scale, but is general and high order. In earthquake science, the hypersingularity has been exactly treated by transferring one derivative off the kernel and onto the slip gradient, exploiting a local reference frame attached to the fault surface that rotates with the geometry and contributes curvature terms \citep[e.g.,][]{sato2020paradox, RomanetOzawa2022}.

The mollification approach here is a direct application of ideas developed in fluid mechanics.  The regularized Stokeslet method \citep{Cortez2001, CortezFauciMedovikov2005} replaces a point force with a smoothed force concentration characterized by a length scale $\eps$, and can be analytically integrated over arbitrary boundaries discretized into triangular elements \citep{FerrantiCortez2024}.  Here, we apply this idea to the displacement-discontinuity formulation of elastostatics constructing a mollified elastic dislocation theory (MEDT), an analog of previous closed-form results \citep[e.g.,][]{NikkhooWalter2015} but for faults that are not infinitely thin. In addition to providing non-singular stress and strain fields the analytic integration of each triangle's kernel contribution effectively decouples the regularization parameter from the element discretization scale \citep{FerrantiCortez2024}.  The closed-form integrals are computed using Claude Code, and we demonstrate how these elements can be used to construct straightforward collocation boundary element solutions for Earth models with topography and inhomogeneous material properties.

\section{Cortez blob mollification and the integration approach over a triangle}
The idea central to the mollification approach is to remove the possibility of singular solutions by regularizing the solutions so that no denominator can go to zero.  Specifically, we replace the point delta function source of CEDT with a Cortez blob \citep{Cortez2001},

\begin{equation}
  \phi_\eps(r) = \frac{15 \eps^4}{8\pi (r^2+\eps^2)^{7/2}}
\end{equation}

subject to $\int \phi_\eps \dd V = 1$. Note that this spatial smearing of slip has a characteristic length scale $\eps$ and is strongly localized (Figure \ref{fig:fig_blob_slip_cartoon}) but does not have a finite width. With this model, in a strict sense, fault slip occurs not just on a fault surface but everywhere in the medium, albeit with the majority of slip occurring in a narrow region, with a power law decay away from the core.  The motivation here is one of mathematical and kinematic convenience although we note that some fault damage zones may lack well-defined boundaries: microcrack density and off-fault plastic strain fall off continuously with distance from the principal slip surface, roughly as a power law \citep[e.g.,][]{SavageBrodsky2011, MitchellFaulkner2009, FaulknerEtAl2010}.

To keep the presented expressions readable, rather than leaping to the displacement discontinuity case we focus on the Kelvin kernel to make the fundamental ideas clear.  The hypersingular displacement discontinuity kernels, which are of primary interest, are far longer and are shared only as code.  Convolving the elasticity force Green's function with $\phi_\eps$ yields a mollified Kelvin kernel,

\begin{equation}
  G_{ij}^\eps(\ve d)
    = \frac{1}{16\pi\mu(1-\nu)}
      \left[ \frac{(3-4 \nu) \del_{ij}} \Req
           + \frac{d_i d_j}{\Req^3}
           + \frac{2(1-\nu) \eps^2 \del_{ij}}{\Req^3} \right]
\end{equation}

where $\Req=\sqrt{r^2+\eps^2}$.  This satisfies the regularized Navier equations $(\mu\nabla^{2}\del_{ij} + (\lambda+\mu)\partial_i\partial_j) G^{\eps}_{jk} = -\del_{ik} \phi_\eps$, where $\lambda = 2\mu\nu/(1-2\nu)$ is the first Lam\'e parameter. Note that the factor $2(1-\nu)$ in the third term is not a free constant \citep{CortezFauciMedovikov2005}. Instead it is required to satisfy the governing Navier equations and it reduces to $1$ in the incompressible limit $\nu=1/2$, and marks the mollified Kelvin Green's function as distinct from the unmollified case in a way other than the denominators containing $r$.  From this mollified Kelvin kernel we obtain the DD displacement and stress kernels by differentiation. Using the standard elastic stiffness tensor for the isotropic homogeneous case $C_{ijkl}=\lambda\del_{ij}\del_{kl}+\mu(\del_{ik}\del_{jl}+\del_{il}\del_{jk})$, the displacement and stress kernels are $U_{ij} = -\bigl( \mu n_m \partial_m G_{ij} + \lambda n_j \partial_m G_{im} + \mu n_k \partial_j G_{ik} \bigr)$ and $K_{mn,k} = -C_{mnpq} C_{kjrs}n_j \partial_p \partial_r G_{qs}$ respectively.

The integration of the mollified kernels over an arbitrary triangle was done specifically with Claude Code 4.6 and was later adversarially critiqued by Claude Fable.  The resulting closed form expressions are long and not reproduced here but are shared as code.  No new techniques were used in this approach, which is instead characterized by detailed bookkeeping and the application of known methods from the Green's function and boundary element literature.  The core integration approach is based on the fact that every component of the displacement discontinuity kernel reduces to a sum of 1) a geometric solid angle calculation \citep[e.g.,][]{VanOosteromStrackee1983}, 2) line integrals along each side of the triangle \citep[e.g.,][]{yoffe1960angular, ComninouDundurs1975}, and 3) in-plane moment integrals, so that the per-triangle integration is exact for any $\eps>0$. The core result is that the mollification introduces an effective observation height $h_\eps = \sqrt{z^{2}+\eps^{2}}$, with $z$ the signed distance from the observation point to the plane of the triangle, so that the mollification amounts to a small change of variables in the classical per-triangle integrals.

\begin{figure}[htp]
    \centerline{\includegraphics[width=1.00\textwidth]{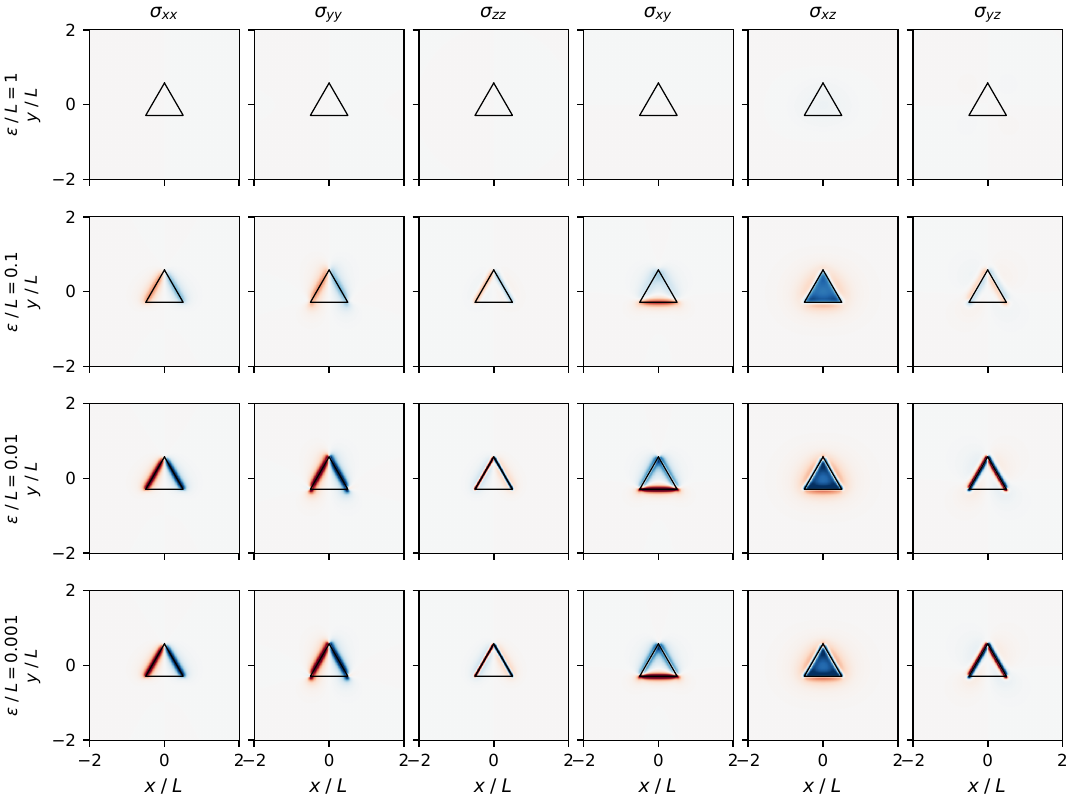}}
    \caption{Mollified analytic displacement discontinuity stress field on the same plane as Figure \ref{fig:fig_triangle_stress_components}. Rows correspond to $\eps/L$ values of 1, 0.1, 0.01, and 0.001. The field converges to a sharp limit as $\eps$ approaches $0$ and remains bounded throughout.}
    \label{fig:fig_eps_sweep_stress_field}
\end{figure}

While closed-form integration of the singular kernel is bounded for any off-plane observation, Gaussian quadrature of the singular kernel produces near-element speckle and blow-up because no fixed-order rule can resolve the $1/r^{3}$ integrand (Figure \ref{fig:fig_triangle_stress_components}). In contrast, the closed-form integration of the mollified kernel is smooth and bounded by construction (Figure \ref{fig:fig_eps_sweep_stress_field}) showing pointwise convergence to a sharp limit while preserving boundedness even at the smallest $\eps$.

The analytic integration of the mollified displacement discontinuity kernel is advantageous not simply because it is non-singular and exact but also because it removes a source of numerical integration error that linked $\eps$ to $h$ \citep{CortezFauciMedovikov2005}. In this classical case, as $h$ decreases $\eps$ must decrease proportionally and, unfortunately, as $\eps$ decreases the quadrature error grows ($\eps^{-3}$), so deliberate choices must be made to simultaneously tune mesh and mollification length scales.  In the Stokes flow case investigations of the $\epsilon{-}h$ scaling associated with the analytic integration of the mollified kernels over triangular elements demonstrated that replacing numerical quadrature with analytic per-triangle integration eliminated the $\epsilon{-}h$ coupling because no numerical quadrature error remains to propagate \citep{FerrantiCortez2024}.

\begin{figure}[htp]
    \centerline{\includegraphics[width=1.00\textwidth]{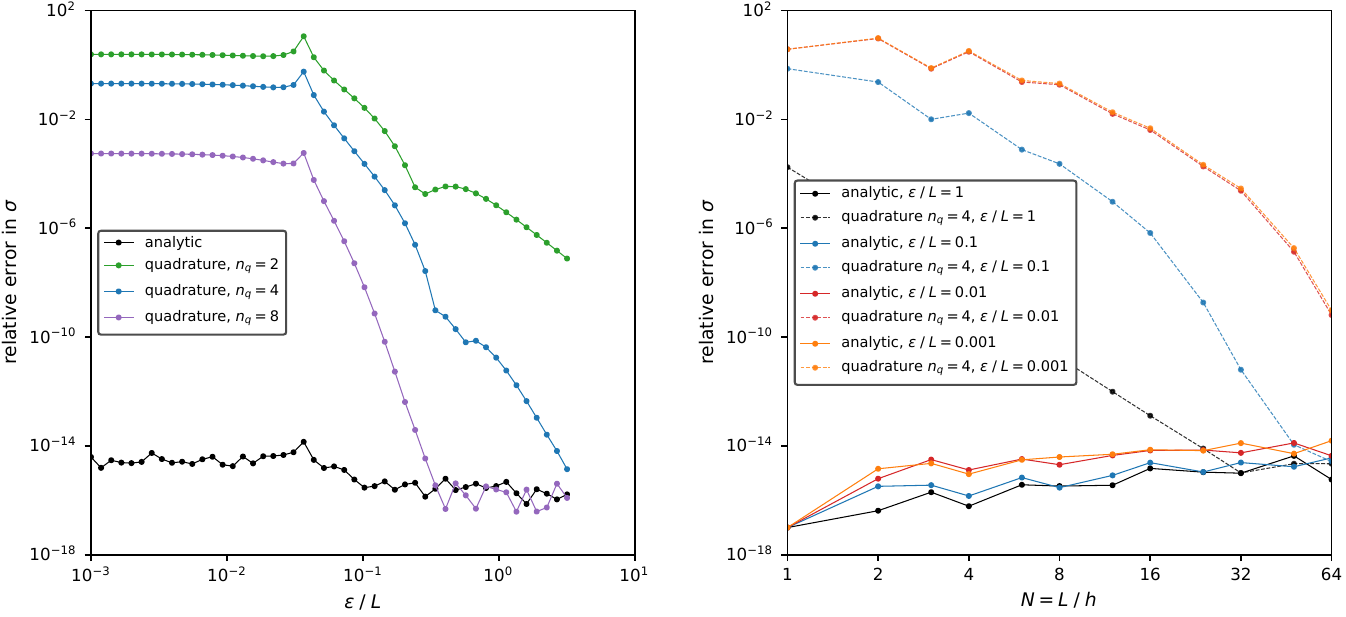}}
    \caption{Decoupling of $\eps$ from $h$ for the mollified DD stress kernel. Subdivision benchmark of a single equilateral source triangle; relative Frobenius error of the recovered $\sigma$ at $(0,0,0.05L)$. Upper panel: fixed $h=L/8$, sweep $\eps$. The analytic curve sits at machine precision; quadrature curves diverge as $\eps$ goes to zero at a rate consistent with the $\eps^{-3}$ scaling of the singular kernel. Lower panel: fixed $\eps$, sweep $N=L/h$. Analytic curves coalesce on a single horizontal line (decoupled); quadrature curves stratify by $\eps$.}
    \label{fig:fig_eps_h_decoupling}
\end{figure}

\begin{figure}[htp]
    \centerline{\includegraphics[width=0.55\textwidth]{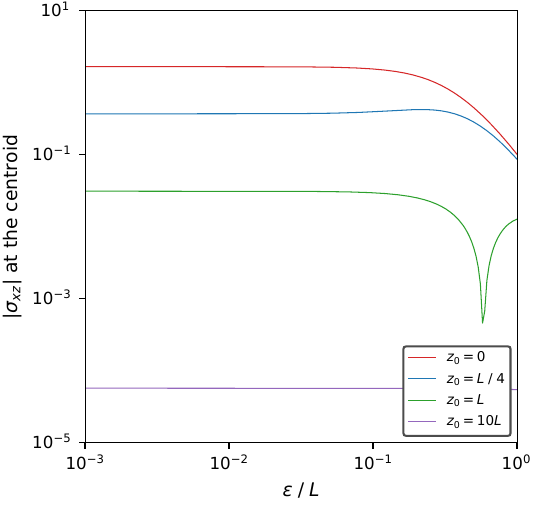}}
    \caption{On-fault elastic stress versus the mollification scale, for the unit-triangle source of Figures \ref{fig:fig_blob_slip_cartoon}--\ref{fig:fig_eps_h_decoupling} (unit edge $L$, unit strike slip $s$, $\mu=1$, $\nu=1/4$), plotted at stand-off distances $z_0 = 0$, $L/4$, $L$, and $10L$ (red, blue, green, purple).}
    \label{fig:fig_onfault_stress_eps_sweep}
\end{figure}

\begin{figure}[htp]
    \centerline{\includegraphics[width=0.55\textwidth]{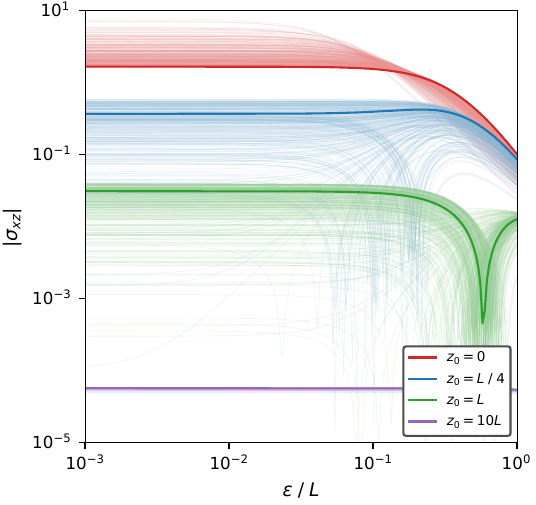}}
    \caption{Sensitivity of stresses to triangle shape for all four stand-off distances of Figure \ref{fig:fig_onfault_stress_eps_sweep}.  The light lines show cases for 300 randomly shaped mollified triangular dislocations of equal area. Vertices are drawn uniformly at random, centroids are at the origin, and areas are rescaled to that of the reference equilateral triangle.  At $z_0 = 10L$ all 300 realizations lie within $13\%$ of the equilateral (the median within $0.5\%$); on the fault plane the spread across shapes is a factor of several.}
    \label{fig:fig_shape_ensemble}
\end{figure}

Through element subdivision experiments we demonstrate that an analog to the Stokes flow decoupling result holds for the elastostatic case (Figure \ref{fig:fig_eps_h_decoupling}).  We start by calculating the exact stress from constant slip $\Du=(1, 0, 0)$ on an equilateral triangle at an observation point just off the fault surface $\obs=(0, 0, 0.05L)$.  We then repeatedly subdivide this reference triangle into sub-triangles and sum per sub-triangle stress contributions both analytically and by Gauss quadrature.  This gives two approximations to the reference triangle's exact stress, whose errors can be compared directly.  Numerical results across a suite of subdivision levels indicate that the analytic error is at machine precision while the numerical quadrature curves depend on $\eps$ (Figure \ref{fig:fig_eps_h_decoupling}).  At a practical level this elastostatic analog of the Stokes result \citep{FerrantiCortez2024} enables us to use $\eps$ values orders of magnitude below the mesh length scale $h$, which makes the boundary element examples below accurate at the few-percent level. Note that for faulting problems involving elastic stress the anelastic eigenstress from the spatially extended fault slip should be subtracted in the Eshelby \citep{Eshelby1957} sense.  This bookkeeping is a pointwise version of the classical source representation where slip is equivalent to a distribution of body forces which CEDT confines to a surface of zero thickness.  These stresses can be calculated on and off the fault plane as a function of $\eps/L$ and show convergence to constant values as $\eps/L$ approaches zero (Figure \ref{fig:fig_onfault_stress_eps_sweep}), demonstrating that the mollification scale may be selected for smoothness, conditioning, or as a physical length scale rather than for accuracy.  Note also that the stress value plateau for low $\eps/L$ values depends on element geometry (Figure \ref{fig:fig_shape_ensemble}) so element-wise on-fault stresses should be interpreted in the context of mesh quality and may motivate higher-order slip representations to mitigate this effect.

\section{Collocation boundary element models with mollified kernels}
\begin{figure}[htp]
    \centerline{\includegraphics[width=0.60\textwidth]{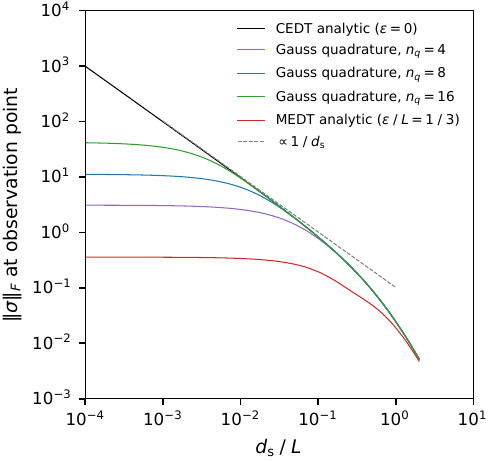}}
    \caption{Near-element stresses for observation points near to a single equilateral source triangle (edge $L=1$) in $z=0$ with unit slip $\Du=(1,0,0)$.  We consider a range of in-plane observation points approaching the triangular source's right vertex, at a distance $d_{\mathrm{s}}$, and plot the Frobenius norm of the integrated stress tensor at the observation point. The closed-form analytic integral of the singular ($\epsilon=0$) kernel of CEDT is finite everywhere but diverges as ${\sim} 1/d_{\mathrm{s}}$ as the observation approaches the source.  Gauss quadrature of the singular kernel at $n_q\in\{4,8,16\}$ saturates at incorrect plateau values. The MEDT analytic ($\eps/L=1/3$) saturates at the smoothed limit and is uniformly bounded.}
    \label{fig:fig_near_element_stress}
\end{figure}

\begin{figure}[htp]
    \centerline{\includegraphics[width=1.00\textwidth]{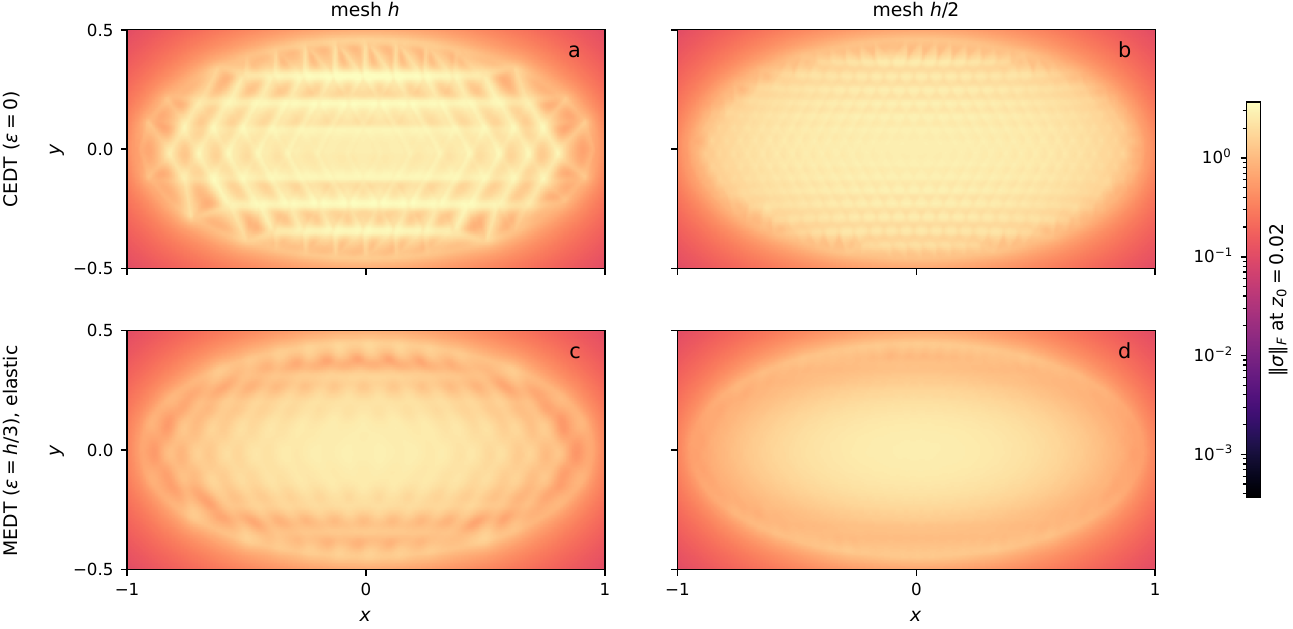}}
    \caption{CEDT (top row) and MEDT (bottom row) calculated stresses on an elliptical crack (semi-axes 0.95 and 0.45, $z=0$ plane, $\mu=1$, $\nu=1/4$) with the smoothly varying slip taper $s=(1-\rho^2)^{3/2}$, where $\rho$ is the normalized elliptical distance from the center, which has zero gradient at the rim. For both the CEDT and MEDT cases two discretizations are used, one with 192 elements (left column) and one with 775 elements (right column).  All maps share one log color scale on the plane $z_0=0.02$ above the crack. Outside the source CEDT and MEDT agree but inside the slipping area, by definition, CEDT is more mesh sensitive than MEDT.}
    \label{fig:fig_elliptical_crack_mesh_sensitivity}
\end{figure}

Collocation boundary element approaches have the merit of simplicity \citep{CrouchStarfield1983} as compared with Galerkin methods \citep[e.g.,][]{SirtoriS.1992Agsb, sutradhar2008symmetric}.  It is worth some discussion of why the hypersingular kernels of CEDT \citep{Okada1992, Meade2007, NikkhooWalter2015} are not well suited for the collocation BEM problem even though they are exact and efficient for forward stress evaluation at finite distances away from fault surfaces.  Note that this is only relevant for elastostatic BEMs with traction boundary conditions (free surfaces, matching boundary conditions at material interfaces, stress constraints) and on-/near-fault stress evaluation. Traction BEM matrix entries are calculated from the stress kernel between a source triangle and an observation point at the collocation site of the source triangle itself and all other triangles.  For ``coincident'' terms on the main diagonal, where the observation sits on the source itself, the $1/r^3$ integral is divergent, requiring a separate analytic limit specific to the source-element geometry.  Similarly for ``near-field'' collocation BEM matrix entries close to the main diagonal, where the observation sits at a distance comparable to the source element length scale ($h_{\text{near}}$), the integrand is finite but singular-like as it scales with $1/h_{\text{near}}^3$ while the integration domain is $O(h^2)$, so any fixed-order quadrature loses several digits per refinement step as the relative error grows at small inter-element distances (Figure \ref{fig:fig_near_element_stress}).  The assembly of a BEM matrix solver built on CEDT singular kernels consequently needs 1) a special self-element formula derived for each element type and boundary-condition combination and 2) adaptive or analytical near-field treatment of the off-diagonal blocks adjacent to the diagonal, and both requirements grow in complexity as soon as slip is non-uniform, the surface is curved, or material interfaces are introduced.  Any fault surface, planar or not, tessellated with constant-slip flat triangles and spatially variable slip over the domain will exhibit jumps along every shared edge wherever adjacent elements carry different slip.  The CEDT stress field develops $1/r$ line singularities at these locations (Figure \ref{fig:fig_elliptical_crack_mesh_sensitivity}) and the CEDT ($\eps=0$) stress field is always a direct image of the triangulation. In contrast, the MEDT elastic field at fixed $\eps=h/3$, computed from an identical tessellation, is an image of the slip distribution and the effects of mesh geometry disappear with mesh refinement (Figure \ref{fig:fig_elliptical_crack_mesh_sensitivity}).

With this rationalization for the potential utility of MEDT in the construction of collocation elastostatic BEM models, we now describe two simple reference problems to demonstrate the approach.  The foundation for collocation BEM is the construction of a matrix that relates boundary conditions to slip on the DD elements that define the boundaries \citep[e.g.,][]{CrouchStarfield1983, BrebbiaTellesWrobel1984, Bonnet1995, Aliabadi2002} and we build this for the second of the two problems.  The first evaluates the mollified kernels directly in a homogeneous half space like geometry, and the second is a collocation BEM featuring inhomogeneous material properties and non-planar topography.

\begin{figure}[htp]
    \centerline{\includegraphics[width=0.85\textwidth]{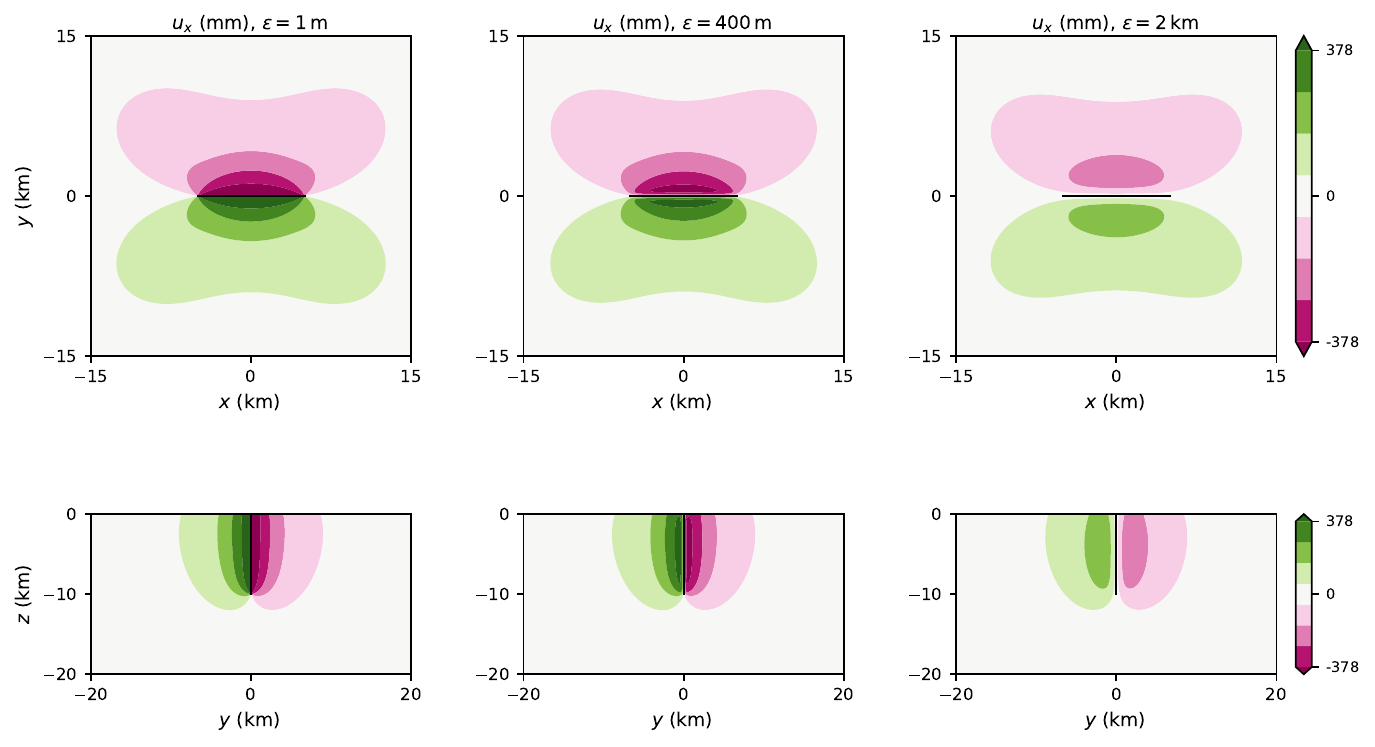}}
    \caption{Map-view and cross-section displacements for the half-space reference problem with 1~m of strike slip on a vertical, surface-breaking rectangular fault (black line), 10~km long and 10~km deep, for mollification scales $\eps$ of 1~m, 400~m, and 2~km (left to right).  The top row shows the fault parallel displacement $u_x$ on a horizontal plane 2~km below the free surface.  The bottom row shows $u_x$ on a vertical cross-section perpendicular to the fault.  With increasing $\eps$ the displacement maximum migrates away from the fault, and the discontinuity across the fault is replaced by a smooth gradation across the fault core.}
    \label{fig:fig_halfspace_like_displacements}
\end{figure}

The first application is a half space like problem with a single rectangular fault at several $\eps$ values, computed directly from a mollified half-space (approximate Mindlin type) form of the kernels rather than from a boundary element solve.  A single surface-breaking vertical rectangular fault, 10~km long and 10~km deep and constructed from two triangular elements, is parameterized with 1~m of strike slip. We calculate the displacements for characteristic fault widths $\eps$ of 1, 400, and 2,000~m (Figure \ref{fig:fig_halfspace_like_displacements}).  These examples show how with increasing $\eps$ the maximum coseismic displacement migrates away from the center of the faulting region introducing a smooth gradation of displacement across the core of a finite-width fault zone.

The second demonstration is designed to highlight the effects of material heterogeneity and topography.  The model domain extends $\pm200$ km in each direction and the top surface has non-planar topography in some cases. A localized disk-shaped (75~km radius, 50~km depth) region of $10\times$ lower shear modulus is included toward the upper left of a single vertical strike-slip fault.  The surface mesh is conformal with the fault trace for both the planar and non-planar topography cases.  A mollification length scale of $\eps = 3$~km is used throughout.  The BEM formulation for homogeneous but differing material properties is written for each subdomain, and then a global system of equations is formed with matching displacement and traction boundary conditions at the interfaces.  For our model, with an inclusion $\mathrm A$ embedded in a host body $\mathrm B$, displacement and traction continuity are $\ve u_{\mathrm A} = \ve u_{\mathrm B}$ and $\sigma_{\mathrm A} \cdot \normal_{\mathrm A} + \sigma_{\mathrm B} \cdot \normal_{\mathrm B} = 0$.  Note that because $\normal_{\mathrm A} = -\normal_{\mathrm B}$, the BEM matrix entries in the host-body rows associated with the interface require a sign flip, while the inclusion's rows do not.

\begin{figure}[htp]
    \centerline{\includegraphics[width=0.85\textwidth]{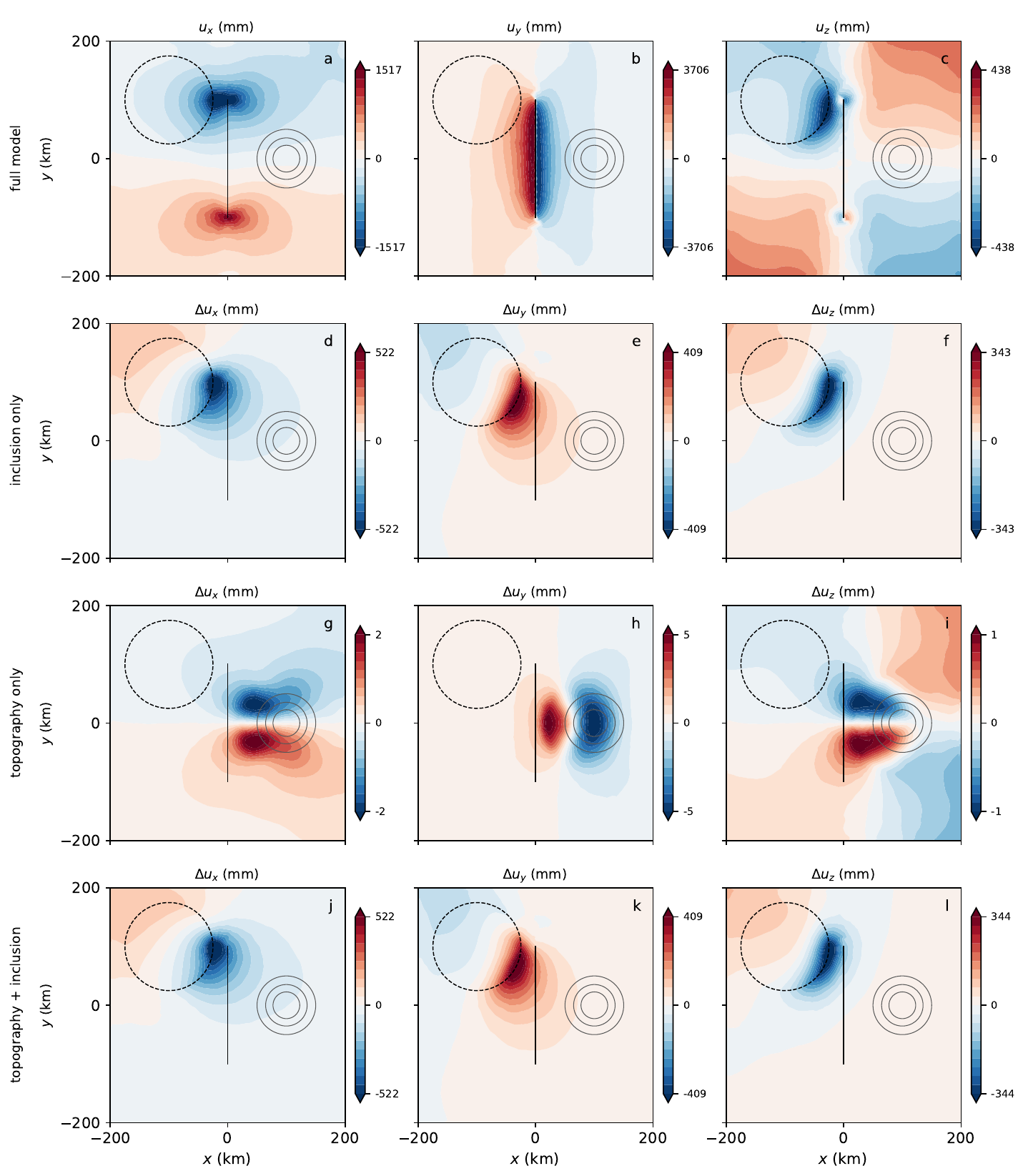}}
    \caption{Example effects of topography and material heterogeneity in a model featuring a vertical strike-slip fault (black line) $x=0$, $|y| \le 100$~km, extending from the surface to a depth of 20~km with 10~m of strike slip.  The lower modulus inclusion (dashed circle) has a radius of 75~km, a depth of 50~km, and a shear modulus $\mu_i=\mu/10$. Synthetic topography (gray contours) is represented as a Gaussian hill with a maximum elevation of 2~km, a width parameter of $30$~km, and is centered 100~km east of the fault. Columns are the surface displacement components $u_x$, $u_y$, $u_z$.  The top row shows displacements from a model with topography and inclusion. The lower three rows show the differential displacements relative to a homogeneous model without topography.  The second row shows the inclusion effect alone (perturbation $<0.8$~m), the third row shows the topography effect ($<6$~mm), and the fourth row shows the combined perturbation, which is not visually distinguishable from the case without topography.}
    \label{fig:fig_topography_inclusion}
\end{figure}

We consider the influence of topography and of the material inclusion, describing each relative to the case with a flat top surface and no inclusion (Figure \ref{fig:fig_topography_inclusion}).  Adding the lower-modulus inclusion perturbs the surface displacement by up to $0.8$~m near the weaker region, even though its center lies 100~km from the fault.  The topographic effect (2~km of relief) is three orders of magnitude smaller than the coseismic signal and two orders of magnitude smaller than the inclusion effect, and is spatially localized near the synthetic topographic hill.  The joint (inclusion and topography) perturbation case is visually indistinguishable from the inclusion-only case, indicating that topographic relief of a few kilometers located a rupture length away from the fault is a negligible modifier of coseismic surface displacement compared with an order of magnitude modulus contrast at the same distance.

Several limitations of the present implementation deserve note.  The BEM uses a far-field-truncating box rather than half-space image kernels, so the accuracy near the box boundary is fixed by the box size; we mitigate this by placing the box boundaries far from the fault ($\pm 200$~km laterally and $200$~km deep in the showcase of Figure~\ref{fig:fig_topography_inclusion}), ample for the 200~km fault.  The direct displacement BEM on a closed bounded soft inclusion has a three-dimensional rigid-body translation nullspace, which we suppress with a single centroid pinning element, and more critically the $90^\circ$ corner where the cylinder's top disk meets its side wall produces a localized singularity that may be ameliorated with more robust BEM formulations \citep[e.g.,][]{BurtonMiller1971} or possibly minimized with higher order shape functions as previously suggested \citep{FerrantiCortez2024}.

\section{Conclusions}
We have described and shown examples from a mollified elastic dislocation theory that mitigates two challenges facing the application of CEDT to faulting problems.  The extended width of fault zones is represented implicitly through the use of mollified Green's functions that distribute slip spatially, and these same kernels are singularity free by construction.  The interesting question that MEDT prompts is how to choose $\eps$. The $\epsilon{-}h$ decoupling, the elastostatic analog of the motivating viscous flow result \citep{FerrantiCortez2024}, allows us to consider this in the context of accuracy and physical length scales.  From an accuracy perspective this means that in practice $\eps$ should be smaller than the smallest geometric scale of every closed sub-domain, not merely the smallest scale of the field one wants to resolve.

Observations of coseismic faulting may offer a physical rationale for choosing $\eps$.  Geodetic measurements provide observations that have been interpreted as zones of distributed coseismic surface deformation, with mean widths of 154~m and 121~m for the Landers and Hector Mine ruptures \citep{Milliner2015, Milliner2016}, 100 to 200~m for the 2013 Balochistan rupture \citep{Cheng2021} along with a hanging-wall extension zone up to 1~km wide \citep{Vallage2015, Gold2015}, ${\sim}250$~m for the 2016 Kumamoto rupture \citep{Scott2018}, a median of ${\sim}30$~m for localized inelastic failure along the 2019 Ridgecrest ruptures \citep{Barnhart2020, Antoine2021, Xu2020}, and a mean of 835~m for the 2021 Maduo rupture \citep{Li2023}.  There are also a few select observations that have been interpreted as indicative of distributed deformation extending a kilometer \citep{Fialko2002, Cochran2009} or more \citep{RodriguezPadilla2022, Liu2025}.  Because two thirds of the slip in the Cortez blob is accommodated within a zone of width $1.1\eps$ centered on the fault plane, observed deformation zone width might map fairly directly to an effective $\eps$ value, and the observations above suggest values from tens of meters to about a kilometer, with the power-law tail of the blob an approximate stand-in for the slowly decaying damage beyond it. Again, as $\eps$ decouples from $h$, these effective fault-width values might be adopted directly in MEDT to represent the kinematic width of a fault.  In summary, MEDT allows for singularity-free BEM formulations, with no self-element or near-field special cases, that simultaneously represent spatially distributed faulting, topography, and material property variations.

\clearpage
\section*{Data Availability}
Python code to reproduce all figures is available at \url{https://github.com/brendanjmeade/medt_paper} and preserved at: \texttt{DOI: 10.5281/zenodo.22712173}.

\clearpage
\section*{Acknowledgements}
This work has benefitted from conversations with Petros Koumoutsakos, Rishav Mallick, and Ben Thompson.  Claude Code was used not only for the analytic integration but also for the generation of most figures and the construction of the BEM matrix.

\clearpage
\bibliographystyle{plainnat}
\bibliography{refs}

\clearpage

\clearpage

\clearpage

\clearpage

\clearpage

\clearpage

\clearpage

\clearpage

\clearpage

\clearpage
\end{document}